\documentclass[sn-mathphys]{ancillary/sn-jnl}

\usepackage{amsmath,amssymb,amstext}
\usepackage{bookmark}
\usepackage[T1]{fontenc}
\usepackage{graphicx}
\usepackage{natbib}
\setcitestyle{numbers}  

\definecolor{ForestGreen}{HTML}{2e8b21}
\definecolor{NatureRed}{HTML}{f54e42}
\definecolor{NatureBlue}{HTML}{4254f5}
\definecolor{NatureGreen}{HTML}{138f57}
\definecolor{NatureMagenta}{HTML}{6b1345}
\definecolor{NaturePurple}{HTML}{c227b0}
\definecolor{NatureCyan}{HTML}{2b807b}
\definecolor{NatureOrange}{HTML}{f5a142}
\usepackage{hyperref}
\hypersetup{
    colorlinks=true,
    linkcolor=blue,
    citecolor=blue,
    filecolor=magenta,
    urlcolor=cyan,
    }
\usepackage{subcaption}
\usepackage{txfonts}
\usepackage{textgreek}
\usepackage{xargs}
\usepackage{xcolor}
\usepackage{xspace}
\jyear{2025}

\usepackage{etoolbox}
\makeatletter
\newcommand\sendemail[3]{
\edef\@tempa{mailto:#1?subject=#2 }%
\edef\@tempb{\expandafter\html@spaces\@tempa\@empty}%
\href{\@tempb}{#3}}

\catcode\%=11
\def\html@spaces#1 #2{#1
\catcode\%=14
\makeatother

\newcommand{\citationneeded}{\textcolor{ForestGreen}{$^{\rm citation\;needed}$}}
\let\oldtextsigma\textsigma
\renewcommand{\textsigma}{\oldtextsigma\xspace}
\let\oldAA\AA
\renewcommand{\AA}{\text{\oldAA}\xspace}
\let\oldtextdegree\textdegree
\renewcommand{\textdegree}{\oldtextdegree\xspace}
\def\w80{\ensuremath{w_{80}}\xspace}

\newcommand{\kms}{\ensuremath{\mathrm{km\,s^{-1}}}\xspace}
\newcommand{\MSun}{\ensuremath{{\rm M}_\odot}\xspace}
\newcommand{\yr}{\ensuremath{{\rm yr}}\xspace}
\newcommand{\Myr}{\ensuremath{{\rm Myr}}\xspace}
\newcommand{\Gyr}{\ensuremath{{\rm Gyr}}\xspace}
\newcommand{\peryr}{\ensuremath{{\rm yr^{-1}}}\xspace}
\newcommand{\Lsun}{\hbox{\,${\rm L}_\odot$}}
\newcommand{\mum}{\text{\textmu m}\xspace}
\newcommand{\kpc}{\text{kpc}\xspace}
\newcommand{\ZH}{\text{[Z/H]}\xspace}

\newcommandx{\lambdar}[2][1=R,2=]{\ensuremath{\lambda_{\rm {#1}}{#2}}\xspace}
\newcommand{\eps}{\ensuremath{\epsilon}\xspace}
\newcommand{\Mstar}{\ensuremath{M_\star}\xspace}
\newcommand{\Mdyn}{\ensuremath{M_\mathrm{dyn}}\xspace}
\newcommand{\re}{\ensuremath{R_\mathrm{e}}\xspace}
\newcommand{\vstar}{\ensuremath{v_\star}\xspace}
\newcommand{\vnai}{\ensuremath{v_{\NaI}}\xspace}
\newcommand{\sigmastar}{\ensuremath{\sigma_\star}\xspace}
\newcommand{\sigmaestar}{\ensuremath{\sigma_{\star,\mathrm{e}}}\xspace}
\newcommand{\vperc}[1]{\ensuremath{v_{#1}}\xspace}

\newcommand{\vesc}{\ensuremath{v_\mathrm{esc}}\xspace}
\newcommand{\nelec}{\ensuremath{n_\mathrm{e}}\xspace}
\newcommand{\Rout}{\ensuremath{R_\mathrm{out}}\xspace}
\newcommand{\vout}{\ensuremath{v_\mathrm{out}}\xspace}
\newcommandx{\Mout}[2][1=,2=]{\ensuremath{M_{\mathrm{out}{#2}}^{#1}}\xspace}
\newcommandx{\Mdotout}[2][1=,2=]{\ensuremath{\dot{M}_{\mathrm{out}{#2}}^{#1}}\xspace}

\newcommandx{\fluxdcgs}[1][1=-20]{$\times 10^{[#1]}$~erg~s$^{-1}$~cm$^{-2}$~\AA$^{-1}$\xspace}
\newcommandx{\powercgs}[1][1=44]{$\times 10^{#1}$~erg~s$^{-1}$\xspace}
\newcommand{\Av}{\ensuremath{A_V}\xspace}

\newcommand{\hda}{\ensuremath{\mathrm{H\text{\textdelta}_A}}\xspace}
\newcommand{\hga}{\ensuremath{\mathrm{H\text{\textgamma}_A}}\xspace}
\newcommand{\Halpha}{\text{H\textalpha}\xspace}
\newcommand{\Hbeta}{\text{H\textbeta}\xspace}
\newcommand{\Hgamma}{\text{H\textgamma}\xspace}
\newcommand{\Hdelta}{\text{H\textdelta}\xspace}
\newcommand{\Pabeta}{\text{Pa\textbeta}\xspace}
\newcommand{\Hepsilon}{\text{H\textepsilon}\xspace}
\newcommandx{\permittedEL}[6][1=O,2=III,3=,4=,5=,6=]{\text{{#1}\,{\sc {#2}}{#3}{#4}{#5}{#6}}\xspace}
\newcommandx{\semiforbiddenEL}[6][1=O,2=III,3=,4=,5=,6=]{\text{{#1}\,{\sc{#2}}]{#3}{#4}{#5}{#6}}\xspace}
\newcommandx{\forbiddenEL}[6][1=O,2=III,3=,4=,5=,6=]{\text{[{#1}\,{\sc{#2}}]{#3}{#4}{#5}{#6}}\xspace}

\newcommand{\CaII}{\permittedEL[Ca][ii]}
\newcommand{\OIII}{\forbiddenEL[O][iii]}
\newcommandx{\OIIIL}[1][1=5007]{\forbiddenEL[O][iii][\textlambda][#1]}
\newcommand{\OIIIall}{\forbiddenEL[O][iii][\textlambda][\textlambda][4960,][5008]}
\newcommandx{\NIL}{\forbiddenEL[N][i][\textlambda][5201]}
\newcommand{\OI}{\forbiddenEL[O][i]}
\newcommand{\OIall}{\forbiddenEL[O][i][\textlambda][\textlambda][6302,][6366]}
\newcommand{\HeI}{\permittedEL[He][i]}
\newcommand{\NaI}{\permittedEL[Na][i]}
\newcommand{\NII}{\forbiddenEL[N][ii]}
\newcommandx{\NIIL}[1][1=6584]{\forbiddenEL[N][ii][\textlambda][#1]}
\newcommand{\NIIall}{\forbiddenEL[N][ii][\textlambda][\textlambda][6550,][6585]}
\newcommand{\SII}{\forbiddenEL[S][ii]}
\newcommand{\SIIall}{\forbiddenEL[S][ii][\textlambda][\textlambda][6718,][6733]}

\newcommandx{\target}[1][1=]{\text{GS-10578{#1}}\xspace}

\newcommand{\jwst}{\textit{JWST}\xspace}
\newcommand{\hst}{\textit{HST}\xspace}
\newcommand{\ppxf}{{\sc ppxf}\xspace}

\newcommand{\Mdynvalue}{$\Mdyn = 2.0\pm0.5 \times 10^{11}$~\MSun}

\raggedbottom
\begin{document}

\title[The Double-Episode Jet Genesis of eROSITA and Fermi Bubbles]{The Double-Episode Jet Genesis of the eROSITA and Fermi Bubbles}
\author[1,2]{\fnm{Ruiyu} \sur{Zhang}}
\author*[3,4,5]{\fnm{Fulai} \sur{Guo}}\email{fulai@shao.ac.cn}
\author[3,4]{\fnm{Shaokun} \sur{Xie}}
\author[6]{\fnm{Ruofei} \sur{Zhang}}
\author[7]{\fnm{Hanfeng} \sur{Song}}
\author[8]{\fnm{Shumin} \sur{Wang}}
\author[9]{\fnm{Guobin} \sur{Mou}}
\author[1,2]{\fnm{Xiaodong} \sur{Duan}}

\affil[1]{School of Physics, Henan Normal University, Xinxiang 453000, China}
\affil[2]{Center for Theoretical Physics, Henan Normal University, Xinxiang 453000, China}
\affil[3]{Shanghai Astronomical Observatory, Chinese Academy of Sciences, Shanghai 200030, China}
\affil[4]{University of Chinese Academy of Sciences,  Beijing  100049, China}
\affil[5]{Tianfu Cosmic Ray Research Center, Chengdu 610000, Sichuan, China}
\affil[6]{School of Astronomy and Space Science, University of Chinese Academy of Sciences,  Beijing  101408, China}
\affil[7]{College of Physics, Guizhou University, Guiyang 550025, China}
\affil[8]{School of Mathematics and Physics, Handan University, Handan 056000, China}
\affil[9]{School of Physics and Technology, Nanjing Normal University, Nanjing 210023, China}


\abstract{
The  Fermi and eROSITA bubbles are giant gamma-ray and X-ray lobes in the Milky Way, extending up to $\sim$50° and ~$\sim$80° in galactic latitude, respectively, yet their origins remain debated. Using three-dimensional magnetohydrodynamic simulations, we investigate a scenario in which two temporally separated episodes of active galactic nucleus (AGN) jets launched from the Galactic center produce the bubbles, with each structure bounded by a forward shock. Our simulations reveal that the first jet pair, launched 15 Myr ago, forms the outer eROSITA bubbles (extending to $\sim$18 kpc), while the second, launched 5 Myr ago, creates the nested Fermi bubbles ($\sim$10 kpc height). This model broadly reproduces the observed elongated morphology, multi-band X-ray surface brightness distribution, O VIII/O VII line ratios, radio ridge structures, and gamma-ray emissions of the bubbles. Cosmic-ray electrons are accelerated \textit{in situ} at the shock fronts, explaining the sharp edges and nearly uniform gamma-ray surface brightness distribution of Fermi bubbles. The results suggest that the eROSITA and Fermi bubbles encode a time-resolved record of episodic AGN  activity in the Galactic center, providing a physically motivated framework for interpreting their multi-wavelength properties.
}

\maketitle{\bf\large{Introduction}}
\\

The discovery of two colossal structures in the Milky Way, the Fermi and eROSITA bubbles, has sparked significant interest and debate regarding their origins. The Fermi bubbles, identified in 2010 through Fermi-LAT gamma-ray observations, exhibit a bilobular shape extending to $\sim50^{\circ}$–$55^{\circ}$ above and below the Galactic plane, characterized by sharp edges, nearly uniform surface brightness, and a hard spectrum with spectral index of $\sim$2 \cite{Su2010,Dobler2010}. A decade later, the eROSITA bubbles, larger counterparts to the Fermi bubbles, were revealed through the eROSITA and MAXI-SSC soft X-ray data, with structures extending to Galactic latitudes of  $\pm 80^{\circ}$ and longitudes of $-60 ^{\circ}$ to $40^{\circ}$ \cite{Predehl2020,Nakahira2020}. Although Fermi bubbles are widely attributed to energetic activity in the Galactic center, their detailed origin remains debated. The nature of eROSITA bubbles remains even more uncertain, with explanations ranging from nearby supernova remnants within $\sim$150 pc of the Sun \cite{Panopoulou2021,Das2020} to Galactic center outflows at a distance of $\sim$8 kpc \cite{Sofue1977,Yang2022,Lallement2023,Liu2024}. The apparent nesting of these two large-scale structures, together with their distinct spatial extents and spectral properties, presents a fundamental challenge for a unified explanation.

Numerous models have been proposed to explain the origin of Fermi and eROSITA bubbles, including single energetic AGN outbursts \cite{Guo2012a,Yang2012,Mondal2022,Mou2014,Zhang2020}, episodic tidal disruption events \cite{Ko2020,Scheffler2025}, and sustained nuclear star-formation–driven outflows \cite{crocker11,Sarkar2015,Gupta2023,Shimoda2024,Sands2025,Heshou2024}. 
A widely adopted hypothesis suggests that both structures were formed by a single energetic event at the Galactic center, with the forward shock marking the edge of the eROSITA bubbles and the contact discontinuity forming the boundary of the Fermi bubbles (e.g.,  \cite{Guo2012,Guo2012a,Sarkar2015,Mou2014,Mou2015,Yang2022}). 

However, growing evidence suggests that a single impulsive event may encounter difficulties in explaining the combined properties of the eROSITA and Fermi bubbles. Hydrodynamic simulations indicate that the sharp edges of Fermi bubbles (as well as the bipolar X-ray structure at the base of Fermi bubbles; \cite{Bland-Hawthorn2003,BlandHawthorn2019}) can be naturally produced by a forward shock driven by a jet-like outflow \cite{Keshet2017,Zhang2020}, but a single event is unlikely to generate two distinct forward shocks that account for both the eROSITA and Fermi bubbles. In addition, gas densities at radii of $r\sim50$–90 kpc substantially exceed expectations from halo gas models constrained by Fermi bubbles, implying the existence of older nuclear activities \cite{Zhang2021}. The inferred dynamical age of the eROSITA bubbles, $\sim15$–20 Myr, significantly exceeds the $\sim4$–10 Myr age of Fermi bubbles constrained by the dynamics of high-velocity clouds and the cooling timescale of GeV–TeV cosmic-ray electrons in leptonic scenarios \cite{DiTeodoro2020,Ashley2022}, instead, they raise the possibility that 
the eROSITA bubbles and the Fermi bubbles originated from two distinct outbursts\cite{Mou2023}. Together, these constraints point to a temporal mismatch that is difficult to reconcile within single-episode models.

In this study, we explore the scenario in which the eROSITA and Fermi bubbles were formed by two temporally-separated eruptions from the Galactic center, each producing forward shocks that define their respective boundaries. Here, we model these eruptions as AGN-driven jets launched along the Galactic rotation axis, perpendicular to the Galactic plane. Although jet-driven models have been proposed \cite{Zhang2021,Yang2022}, detailed simulations of a double-episode jet scenario remain unexplored. Previous models typically produce only one single forward shock structure, which does not explain two nesting pairs of bubbles (forward shocks) with distinct thermodynamic and radiative signatures. Using three-dimensional magnetohydrodynamic simulations, here we aim to reproduce the morphology, thermal properties, and radio, X-ray, and gamma-ray emissions of the bubbles, providing new insights into their origins, the dynamic history of the Galactic center, and the processes driving Galactic-scale outflows.

\section*{{Results}}

Here we present results from our fiducial simulation, which best reproduces the observed morphology and multi-band X-ray surface brightness of the eROSITA and Fermi bubbles (see Supplementary Information for a parameter study). The corresponding emission maps in the gamma-ray and radio bands are also calculated for direct comparison with observations. In this model, two pairs of kinetic-energy-dominated jets were successively launched at the Galactic center (see Methods). The first jet pair injected a total energy of $E_{\rm total}=3.46 \times 10^{55}$ erg and was followed by the second jet pair 10 Myr later with $E_{\rm total}=1.10 \times 10^{55}$ erg. Both pairs of jets lasted for a duration of $t_{\rm jet}=1$ Myr.

\begin{figure}[htbp]
\centering
\begin{subfigure}{0.45\textwidth}
\centering
\includegraphics[width=\linewidth]{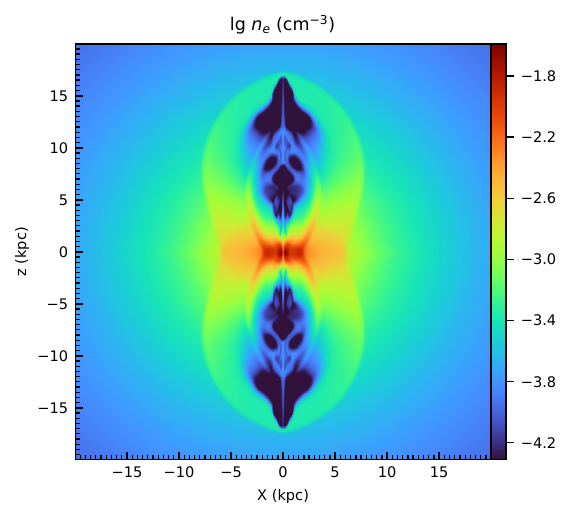}
\caption{}
\label{density16}
\end{subfigure}\hfill
\begin{subfigure}{0.45\textwidth}
\centering
\includegraphics[width=\linewidth]{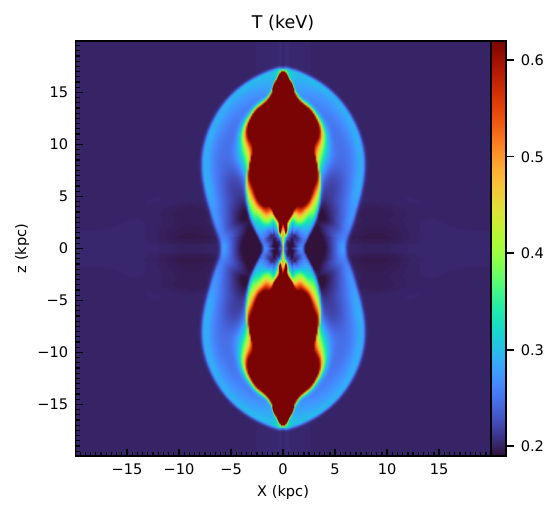}
\caption{}
\label{temperature16}
\end{subfigure}
\begin{subfigure}{0.45\textwidth}
\centering
\includegraphics[width=\linewidth]{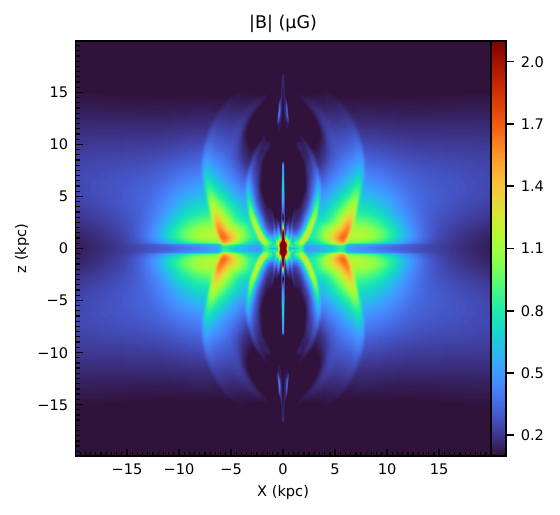}
\caption{}
\label{magnetic15}
\end{subfigure}\hfill
\centering
\begin{subfigure}{0.45\textwidth}
\centering
\includegraphics[width=\linewidth]{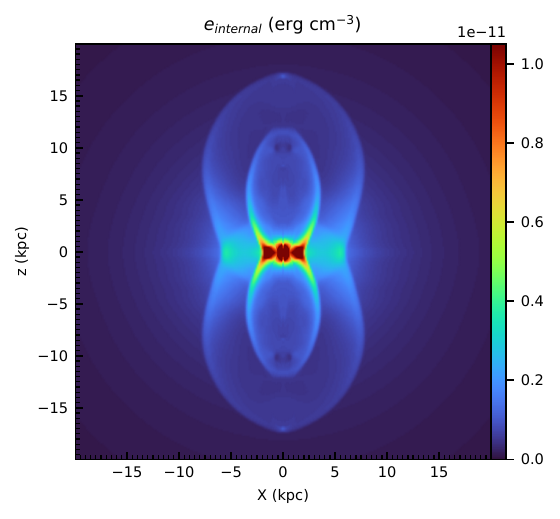}
\caption{}
\label{u_e}
\end{subfigure}\hfill

\caption{Central slices of electron number density (a), temperature (b), magnetic field strength (c), and internal energy density (d) at t = 15 Myr in the fiducial run, where two temporally-separated pairs of AGN jets are launched at t = 0 Myr and 10 Myr, respectively, both lasting for a duration of 1 Myr. Two layers of forward shocks are triggered successively and clearly seen in this figure.}
\label{denT}
\end{figure}

The simulation reveals a nested, shock-driven bubble structure in the Galactic halo. At $t=15$ Myr, two distinct forward shocks are present: an outer shock extending to $\sim18$ kpc above the Galactic plane that corresponds to the eROSITA bubbles (with the northern bubble slightly smaller at $\sim$15 kpc), and an inner shock reaching $\sim10$ kpc that delineates the Fermi bubbles (see Fig.~\ref{denT}).  These shocks define two spatially and dynamically distinct bubble systems with sharp boundaries, naturally reproducing their observed size and elongated morphology.

The temperature structure of the shocked gas reflects this two-stage evolution history. As shown in Fig.~\ref{temperature16}, the gas downstream of the first (outer) shock is compressed and heated to $\sim 0.25-0.3$ keV, forming a broad shell with density gradually decreasing with height from the Galactic plane. Behind this shock, the gas cools away from the shock front as a result of adiabatic expansion. The second (inner) shock then re-compresses this post-shock gas and creates a denser and narrower shell, within which the gas is reheated to $\sim 0.3-0.4$ keV, increasing toward higher latitudes. The internal energy density and magnetic field are also enhanced by the shock fronts. In addition to the shocked ambient gas, the two pairs of bubbles also contain low-density ejecta from both jet outbursts with temperatures exceeding 1 keV.

To validate our double-episode jet scenario, we generate synthetic all-sky maps in the ROSAT energy bands (0.11–2.04 keV) and compare them directly with ROSAT observations. The calculations incorporate bremsstrahlung, line emissions, and foreground HI absorption (see Methods).

\begin{figure}[htbp]
\centering
\begin{subfigure}{0.45\textwidth}
\centering
\includegraphics[width=\linewidth]{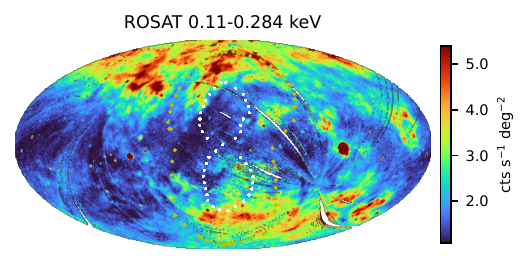}
\caption{ROSAT (R1R2)}
\label{or12}
\end{subfigure}\hfill
\begin{subfigure}{0.45\textwidth}
\centering
\includegraphics[width=\linewidth]{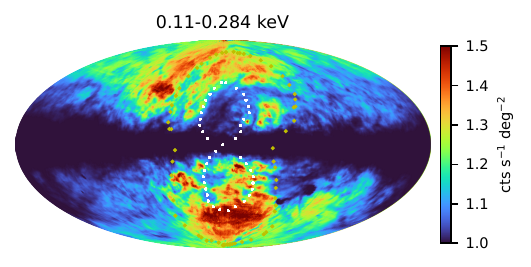}
\caption{Synthetic (R1R2)}
\label{r121}
\end{subfigure}

\vspace{1em} 

\begin{subfigure}{0.45\textwidth}
\centering
\includegraphics[width=\linewidth]{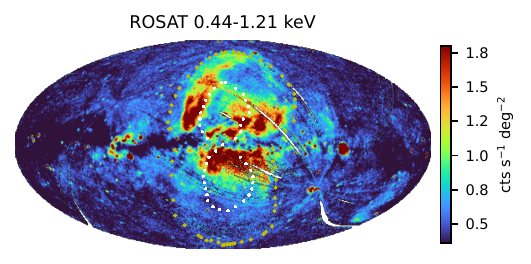}
\caption{ROSAT (R4R5)}
\label{or45}
\end{subfigure}\hfill
\begin{subfigure}{0.45\textwidth}
\centering
\includegraphics[width=\linewidth]{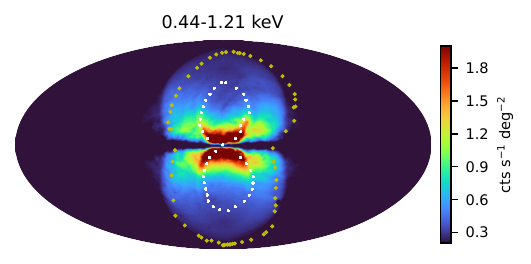}
\caption{Synthetic (R4R5)}
\label{r452}
\end{subfigure}

\vspace{1em} 

\begin{subfigure}{0.45\textwidth}
\centering
\includegraphics[width=\linewidth]{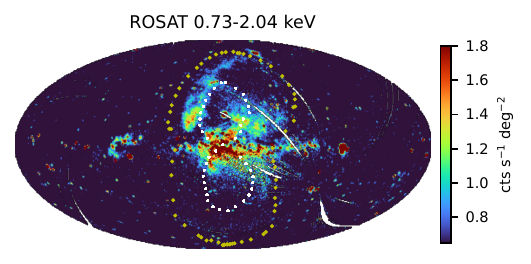}
\caption{ROSAT (R6R7)}
\label{or67}
\end{subfigure}\hfill
\begin{subfigure}{0.45\textwidth}
\centering
\includegraphics[width=\linewidth]{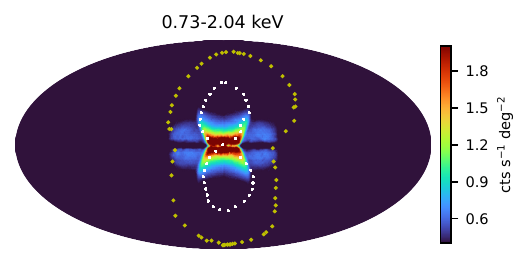}
\caption{Synthetic (R6R7)}
\label{r672}
\end{subfigure}

\caption{Comparison of ROSAT and synthetic X-ray maps in different bands using the Hammer-Aitoff projection at \( t = 15 \, \text{Myr} \). Left column: ROSAT all-sky maps\cite{Snowden1997}; right column: synthetic surface brightness maps. Top row (panels a, b): 0.11--0.284\,keV (R1R2) band, with an additional emission of 1\,cts\,s\(^{-1}\)\,deg\(^{-2}\) added to the synthetic map. Middle row (panels c, d): 0.44--1.21\,keV (R4R5) band. Bottom row (panels e, f): 0.73--2.04\,keV (R6R7) band. Markers: `\( \times \)' for eROSITA bubble edges and `\( + \)' for Fermi bubble edges overlaid on all maps. }
\label{r_combined}
\end{figure}

In the $0.11$ -- $0.284$ keV band (ROSAT R1R2; \cite{Snowden1997}), the X-ray surface brightness drops significantly from high to low latitudes, as shown in the \text{ROSAT} all-sky map (Fig.~\ref{or12}). Due to its large photoionization cross-section in this soft X-ray band, low-latitude ($\vert b\vert < 30^\circ$) emission is strongly absorbed by the intervening HI gas and, therefore, dominated by nearby sources. At high latitudes, the HI column density is about two orders of magnitude lower than in the Galactic plane, allowing X-ray emission from the halo to penetrate. Our synthetic map (Fig.~\ref{r121}) reproduces the observed fading trend from high to low latitudes, with bright patches at high latitudes aligning well with \text{ROSAT} observations. Our synthetic map (Fig.~\ref{r121}) lacks bright low-latitude emission features from nearby supernova remnants and X-ray binaries, such as Vela (\( l, b = 263.9^\circ, -3.3^\circ \), 287\,pc~\cite{Sushch2011}), the Monogem ring (\( l, b = 201.1^\circ, 8.3^\circ \), 300\,pc~\cite{Reich2020}), and Antlia (\( l, b = 276^\circ, 19^\circ \), 250\,pc~\cite{Fesen2021}). These nearby sources remain visible at low latitudes, whereas the eROSITA and Fermi bubbles are significantly attenuated, indicating that the bubbles lie beyond 300 pc, exceeding the Sco-Cen association's distance of 150 pc~\cite{Berkhuijsen1971}. The North Polar Spur (NPS), which fades similarly at low latitudes in the ROSAT map, supports its Galactic-scale origin and is consistent with its association with the eROSITA bubbles. Interestingly, our synthetic surface brightness is typically a few times lower than the \text{ROSAT} data, probably because our model adopts an isothermal halo at $0.2$ keV, while in reality a slightly cooler gas may contribute significantly~\cite{Guo2020}, and the local bubble may also play a role~\cite{Yeung2024}.

At higher energies ($0.44$–$1.21$ keV), the reduced photoelectric absorption reveals the underlying shock-bounded morphology more clearly (Figs.~\ref{or45} and \ref{r452}). Two distinct emission layers emerge, corresponding to the forward shocks driven by the two jet episodes and tracing the boundaries of the eROSITA and Fermi bubbles. Our model predicts enhanced low-latitude emissions due to higher post-shock densities there, forming the bipolar X-ray structure near the Galactic center. Single-episode outburst models, with Fermi bubble edges as a contact discontinuity, fail to explain this X-ray feature. With increasing latitudes, the emission gradually decreases, mirroring the southern hemisphere trend on the \text{ROSAT} map (Figs.~\ref{or45}) and indicating a decrease in the post-shock gas density with latitude.

In the hardest ROSAT band (0.73–2.04 keV; Fig.~\ref{or67}), the same trend of decreasing emission with latitude is seen and, in particular, the eROSITA bubbles become much dimmer than the Fermi bubbles compared to the softer bands. This feature is reproduced in our synthetic map (Fig.~\ref{r672}), and is caused by the lower post-shock gas temperature in the former. From the soft to hard ROSAT X-ray band, the emissivity of the $0.4$ keV gas drops much slower than that of the $0.3$ keV gas. As a result, the central X-shaped structure, i.e., the base of Fermi bubbles, becomes more prominent in this hard band, highlighting how the double-episode jet model shapes the X-ray morphology and spectrum of the eROSITA and Fermi bubbles.

\begin{figure}
  \centering
  \begin{subfigure}{0.3\textwidth}
  \centering
\includegraphics[width=\linewidth]{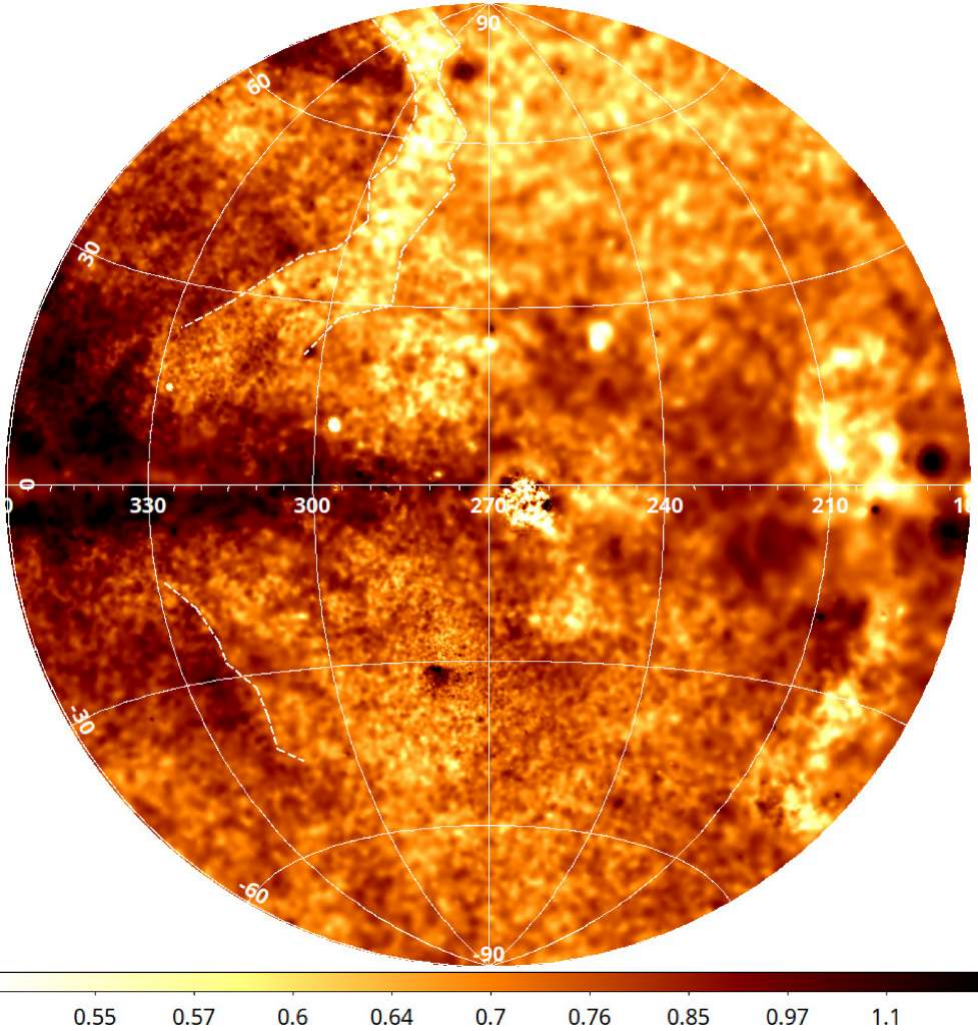}
\caption{Observed}
\label{aafig7}
\end{subfigure}\hfill
   \begin{subfigure}{0.6\textwidth}
   \centering
  \includegraphics[width=\linewidth]{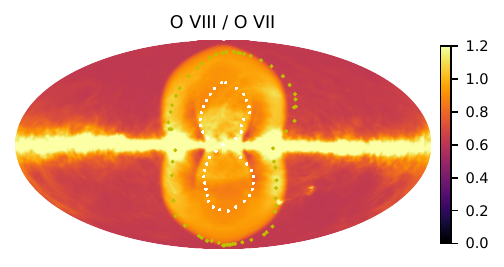}
  \caption{Synthetic}
\label{o7o8}
\end{subfigure}\hfill
  \caption{Comparison of the synthetic map of the line ratio between O VIII (0.614--0.694 keV) and O VII (0.534--0.614 keV) at $t = 15$ Myr in the fiducial run (panel b) with the observed map from the eROSITA data (reproduced with permission \cite{Zheng2024}; panel a).}\label{o7o8compare}
\end{figure}

Independent constraints from X-ray line diagnostics further support our model. The synthetic O VIII/O VII line ratio map generally reproduces the observed enhancement of this line ratio inside the eROSITA bubbles, indicating a shock-heated gas bounded by a mild forward shock (Fig. \ref{o7o8compare}). The inferred shock Mach number ($< 1.5$) and the gradual decrease in the ratio with latitude imply a dynamical age of $\sim15$ Myr, consistent with our model. In single-episode scenarios, adiabatic cooling typically suppresses this ratio near the Galactic center, whereas in our model, the second jet episode reheats the interior of the eROSITA bubbles, maintaining the elevated gas temperature.

\begin{figure}[htbp]
\centering
\includegraphics[width=\linewidth]{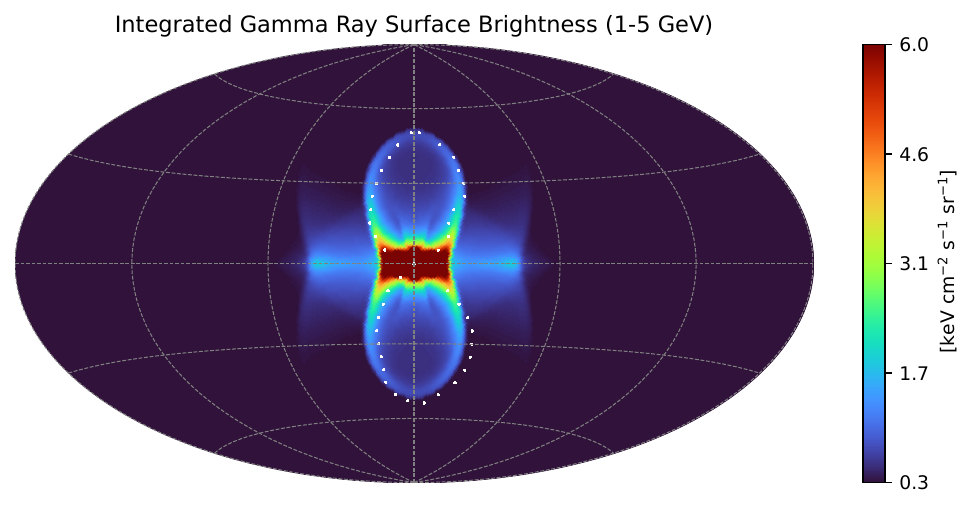}
\caption{Synthetic gamma-ray surface brightness distribution in the 1–5 GeV band at $t = 15$ Myr in the fiducial run. To calculate the gamma ray intenity, we adopt a toy model where the CRe energy density is assumed to be proportional to the thermal energy density of the hot gas, and the gamma-ray emission is modeled through inverse Compton scattering (ICS) of CRes off the interstellar radiation field (ISRF). The spectral indexes of non-thermal electrons are set to $p=2.2$ and $2.4$ inside and outside the Fermi bubbles, respectively. The white dots delineate the outer boundary of the observed Fermi bubbles.}
\label{FB_internal}
\end{figure}

Adopting a toy model for the cosmic ray electron (CRe) distribution (see Methods), we produce the synthetic gamma-ray surface brightness distribution in the 1–5 GeV band (Fig.~\ref{FB_internal}), which reproduces both Fermi bubbles with sharp edges and the potential low-latitude counterpart of eROSITA bubbles. The calculated gamma-ray intensity of Fermi bubbles is roughly consistent with Fermi observations \cite{Ackermann2014}, and the corresponding CRe energy within the bubbles is  $\sim 9.55\times10^{52}$ erg. In our toy model, the CRe energy density is assumed to be proportional to the thermal energy density, resulting in limb brightening at the forward shock in the gamma ray map, which may be alleviated by CRe diffusion toward the bubble interior. Our model highlights regions of enhanced CRe energy density near the edges of the bubbles (Fig.~\ref{u_e}), which is consistent with recent re-analysis of the Fermi-LAT data \cite{Tank2025}, supporting \textit{in situ} cosmic-ray acceleration near the shock fronts.

\begin{figure}[htbp]
\centering
\begin{subfigure}{0.45\textwidth}
\centering
\includegraphics[width=\linewidth]{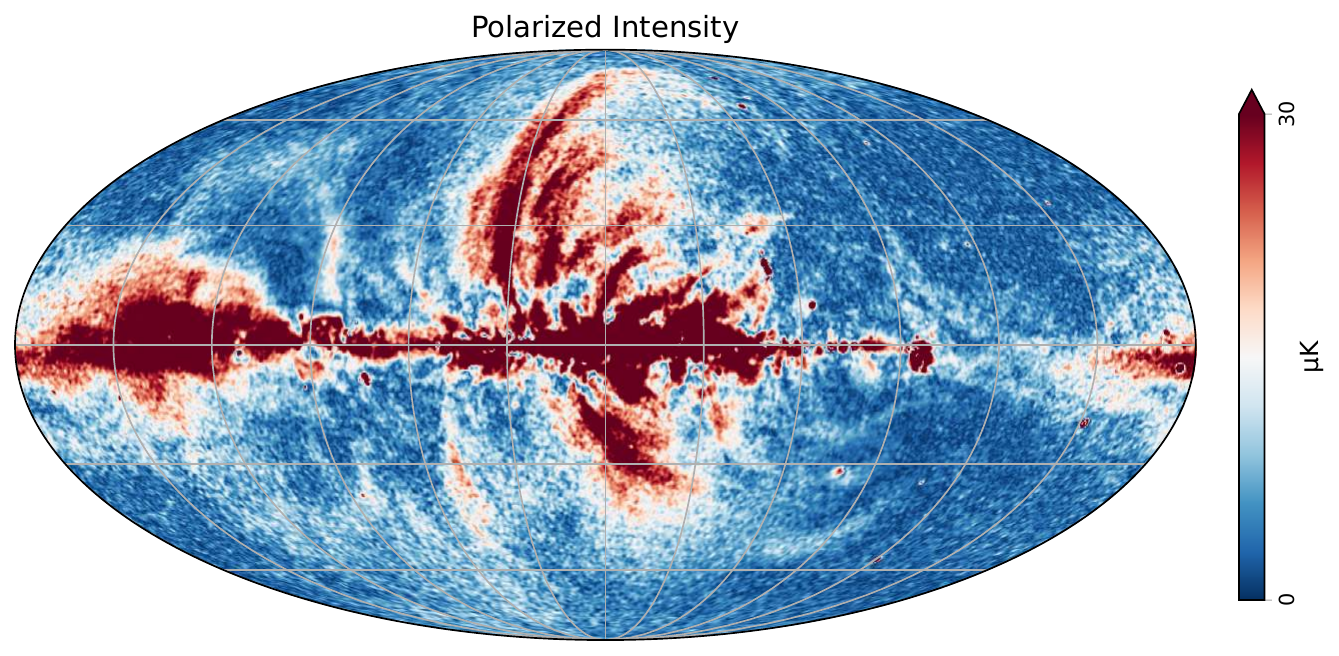}
\caption{Observed}
\label{obs_pi}
\end{subfigure}\hfill
\begin{subfigure}{0.45\textwidth}
\centering
\includegraphics[width=\linewidth]{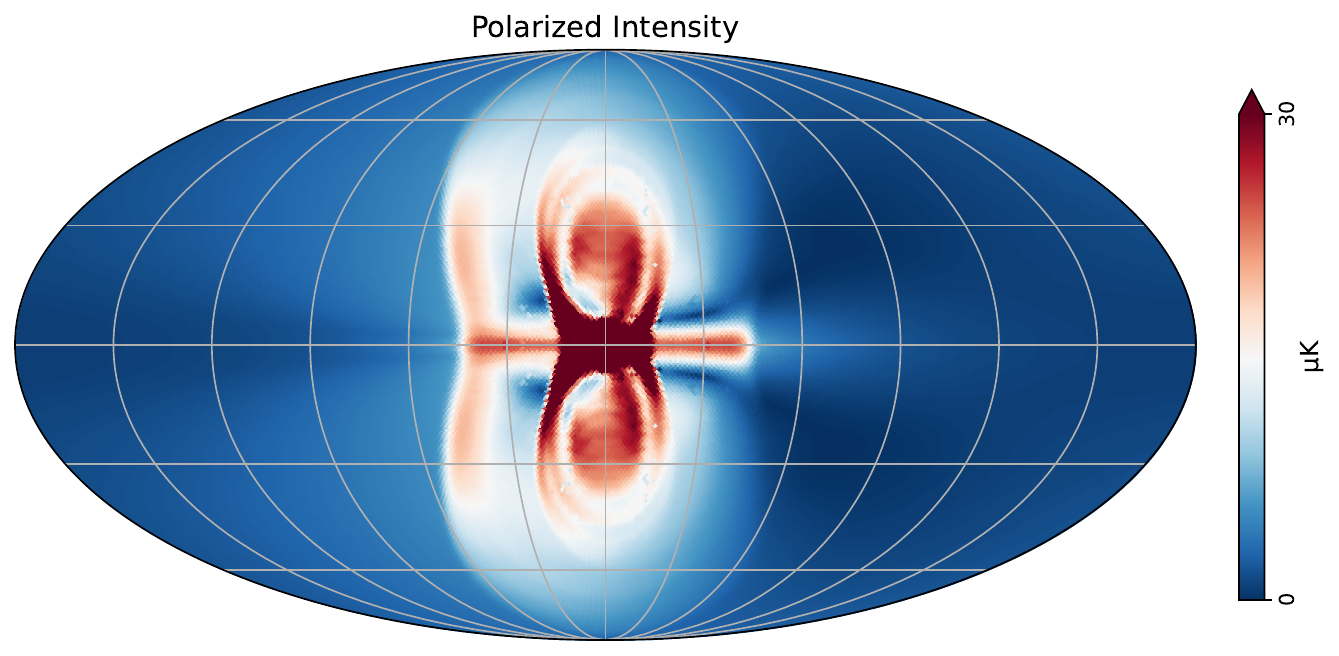}
\caption{Synthetic}
\label{sim_pi}
\end{subfigure}

\vspace{1em}

\begin{subfigure}{0.45\textwidth}
\centering
\includegraphics[width=\linewidth]{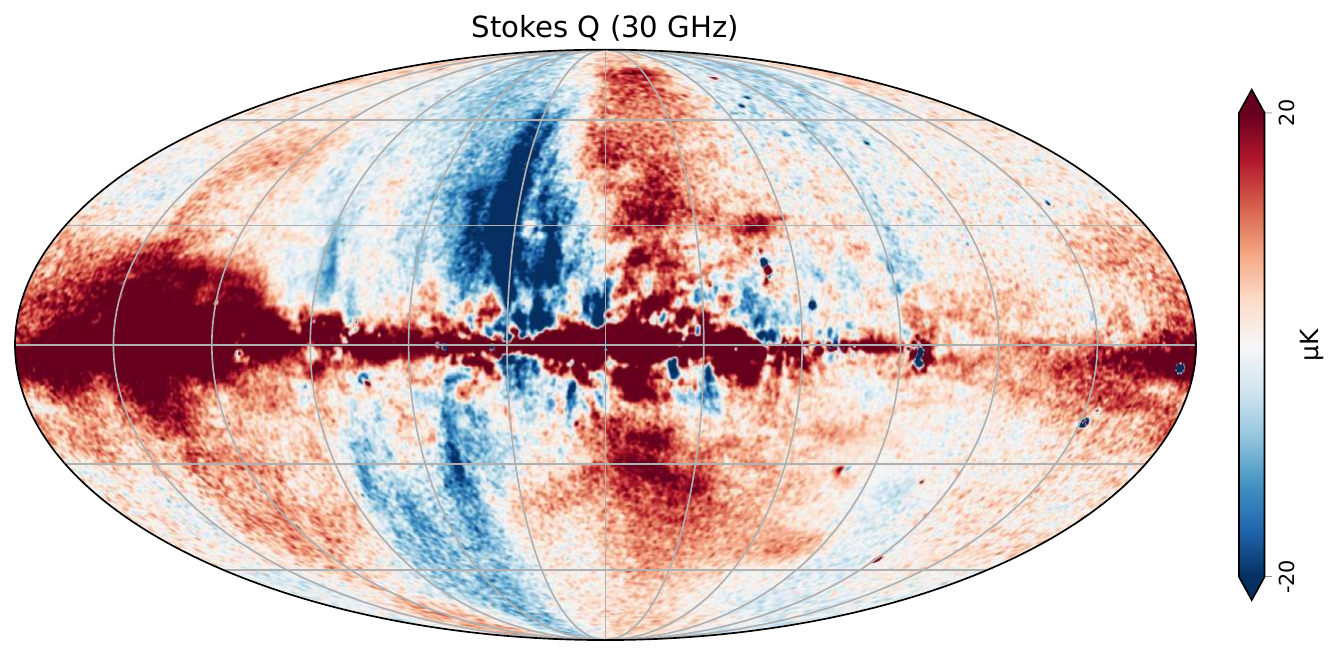}
\caption{Observed}
\label{obs_q}
\end{subfigure}\hfill
\begin{subfigure}{0.45\textwidth}
\centering
\includegraphics[width=\linewidth]{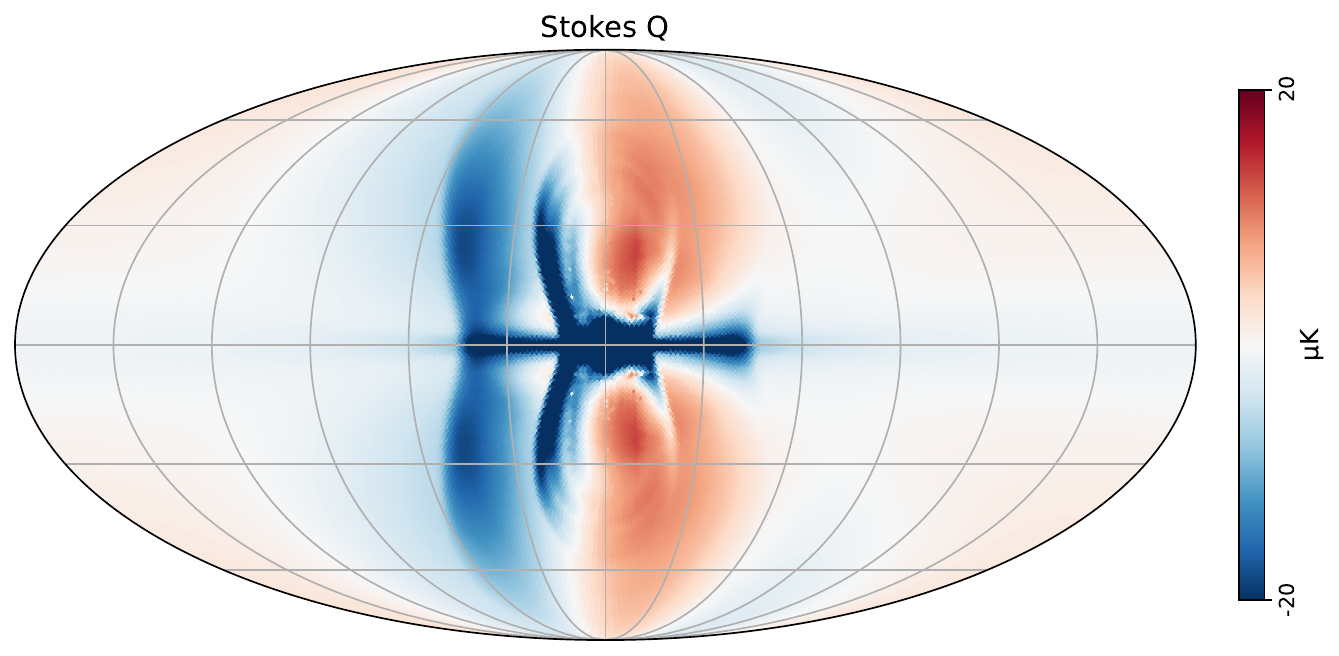}
\caption{Synthetic}
\label{sim_q}
\end{subfigure}

\vspace{1em}

\begin{subfigure}{0.45\textwidth}
\centering
\includegraphics[width=\linewidth]{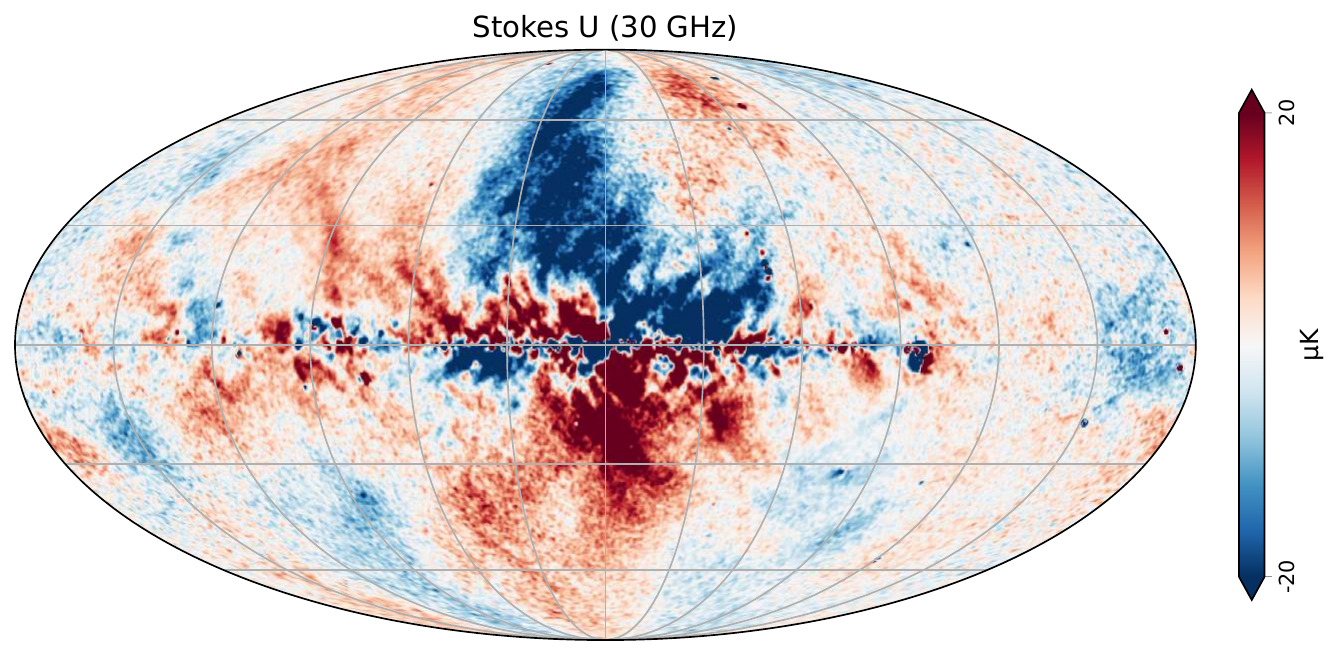}
\caption{Observed}
\label{obs_u}
\end{subfigure}\hfill
\begin{subfigure}{0.45\textwidth}
\centering
\includegraphics[width=\linewidth]{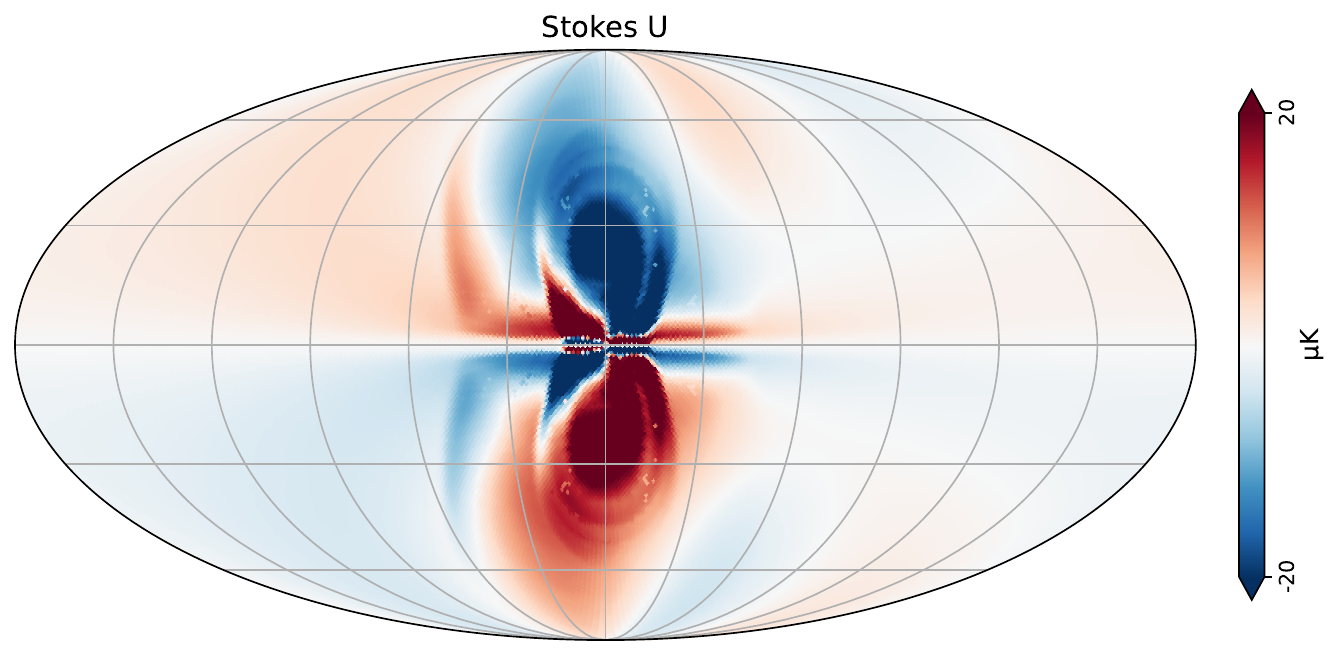}
\caption{Synthetic}
\label{sim_u}
\end{subfigure}

\caption{Polarized intensity (top) and Stokes Q (middle) and U (bottom) polarization maps at 30 GHz. Left column: Observed maps using the WMAP and \textit{Planck} data \cite{Delabrouille2024}; right column: Synthetic maps at $t = 15$ Myr in the fiducial run. As in the gamma-ray calculations, we adopt a toy model where the CRe energy density is assumed to be proportional to the thermal energy density of the hot gas.}
\label{stokes_maps}
\end{figure}

We further computed the polarized synchrotron emission at 30 GHz in our fiducial run to compare with Planck and WMAP observations (Fig.~\ref{stokes_maps}). The polarized intensity map reveals ridge-like structures that align with the shock-compressed bubble edges. The synthetic Stokes Q and U parameter maps broadly reproduce the large-scale polarization features in the observation maps, including the anti-symmetric lobes, which can be explained by the composition of the toroidal and poloidal field components in the inner Galaxy (Fig.~\ref{B0}). These features emerge naturally from the interplay between cosmic-ray electrons and magnetic fields in the simulation, consistent with the observed polarized Galactic synchrotron emission.

\section*{Discussion}


We present a coherent physical scenario in which the eROSITA and Fermi bubbles arise from two temporally separated jet episodes launched from the Galactic center. They drive two successive forward shocks that naturally account for the elongated morphology and sharp outer edges of the bubbles, at $\sim$18 kpc for eROSITA bubbles and $\sim$10 kpc for Fermi bubbles. Remarkably, this model simultaneously reproduces a broad range of observational diagnostics—including multi-band X-ray surface brightness and O VIII/O VII ratios, polarized  synchrotron emission at 30 GHz, and gamma-ray emission. Our results suggest that the bubble structures encode a time-resolved record of episodic AGN feedback in the Milky Way.

A key implication of our model is that the gamma-ray emission of Fermi bubbles is dominated by CRes accelerated \textit{in situ} by the second forward shock and subsequent downstream turbulence, whereas the low-energy aged CRes previously accelerated by the first forward shock may serve as important seed particles. The cooling timescales of the TeV CRes are $\sim$1–2 Myr due to synchrotron and inverse-Compton losses, which is a long-standing challenge for previous ejecta models of Fermi bubbles. Our simulation reproduces both the sharp edges and the gamma-ray surface brightness distribution of Fermi bubbles, supporting an in situ acceleration model for Fermi bubbles as a shock relic of a recent AGN outburst \cite{Zhang2020}.

Our double-episode jet model predicts the existence of multiple populations of CRes, providing a natural explanation for the S-PASS lobes as well. The ejecta of the first jet episode carry CRes that, by the present day, have cooled below gamma-ray-emitting energies but still produce synchrotron emission at radio frequencies. These aged electrons extend beyond the Fermi bubbles toward higher latitudes and are responsible for the S-PASS lobes observed at 2.3 GHz. The second jet episode drives a shock that propagates through the ejecta of the first outburst, accelerating a fresh population of electrons that produce the gamma-ray emission characterizing the Fermi bubbles. A key observational consequence of this two-population model is a spectral break in the radio continuum across the outer edge of the Fermi bubbles: the synchrotron spectrum of the S-PASS lobes should be significantly steeper than that inside the bubbles. This provides a clear, testable signature for future multi-wavelength observations.

When modeling the radio and gamma-ray emissions, we treat the normalizations of the corresponding CRe energy distributions as independent free parameters. The resulting radio normalization exceeds that of the gamma rays by approximately an order of magnitude. If both emissions originated from the same CRe population, matching the observed radio flux would require a magnetic field strength roughly three times higher than that adopted in our simulation ($\sim$3–6 $\mu$G), consistent with estimates in some literature \cite{Shaw2022}. Given the substantial uncertainties regarding whether the radio and gamma-ray emissions trace a single electron population in a system shaped by multiple shock episodes, we adopted independent normalizations as a conservative modeling choice. This approach highlights the need for future multi-wavelength observations to jointly constrain the particle populations and magnetic field structure in the halo.

More broadly, our results suggest that Sgr A* has experienced episodic AGN activity over the past $\sim$15 Myr, with individual jet outbursts reaching kinetic powers of $\sim 10^{41}$–$10^{42}$ erg s$^{-1}$. This would place the Galactic center among a class of low-luminosity yet recurrent AGNs, capable of influencing the nuclear star formation, cosmic-ray injection, and magnetization in the Galactic halo. Independent observational evidence supports this picture on multiple timescales: X-ray echoes from molecular clouds point to a luminous flare roughly 200 years ago \cite{Marin2023}, while the Fermi bubbles and elevated ionization in the Magellanic Stream have been interpreted as signatures of powerful AGN episodes several million years ago \cite{Bland-Hawthorn2013}. More recently, theoretical work has proposed that a merger between Sgr A* and an intermediate-mass black hole around $\sim$10 Myr ago \cite{Cao2025} may have enhanced accretion and subsequently triggered jet activity.

Although alternative scenarios—such as AGN winds, episodic tidal disruption events, or sustained nuclear star formation—may also generate large-scale shocks, models based on continuous or high-frequency energy injection typically produce a single dominant shock structure and face challenges in simultaneously reproducing the sharp double shock fronts, the uniform gamma-ray surface brightness distribution, and the spatial offset between radio and gamma-ray structures. By contrast, the double-episode jet model provides a self-consistent explanation for these features within a unified physical framework. Although our current model does not account for all observed features—such as the enhanced X-ray and radio emission of the NPS—it strongly supports a double-episode jet-driven origin for the Fermi and eROSITA bubbles and highlights the role of low-level, recurrent AGN feedback in shaping the Milky Way's halo.

\section*{{Methods}}

We perform 3D MHD simulations to model the formation of the eROSITA and Fermi bubbles via two successive AGN jet episodes from the Galactic center. Simulations are conducted using the finite-volume code PLUTO v.4.4-patch2 \cite{Mignone2007}, with post-processing to generate synthetic multi-wavelength maps for comparison with observations. The hot gas in the Galactic halo is assumed to be initially isothermal and we neglect radiative cooling, as the cooling timescale of the hot halo gas exceeds the simulation duration by more than an order of magnitude, rendering the cooling process negligible during the whole simulation time.

The MHD equations are solved in a Cartesian coordinate system using the Harten--Lax--van Leer (HLL) Riemann solver \cite{Toro1994}. The $z$ axis aligns with the Galactic rotation axis, and the $x$--$y$ plane corresponds to the Galactic plane. The Sun is positioned at $(-8.3, 0, 0)$ kpc \cite{Abuter2021}. The simulation domain extends from $-30$ to 30 kpc along each axis, with a grid consisting of 128 uniformly spaced zones from $-1$ to 1 kpc and 192 additional uniform zones on each side extending to the outer boundary. Standard outflow boundary conditions are applied at all outer boundaries.

To ensure sufficient line-of-sight integrations for X-ray, radio, and gamma-ray intensity calculations, the gas density profile is pre-computed out to 100 kpc from the Galactic center and the gas densities outside our simulation domain are held fixed during the simulation. Details on initial conditions, jet injection, and post-processing methods are described as follows.

\subsection*{Initial Conditions}

\subsubsection*{Halo Gas Density and Temperature}

The Galactic gravitational potential is modeled as a static multi-component background \cite{McMillan2017}. The hot gas in the Galactic halo is assumed isothermal with temperature $T = 0.2$ keV, consistent with observations \cite{Henley2013,Miller2015,Kataoka2018,Zheng2024,Zhangyi2024}. The density distribution is determined by hydrostatic equilibrium with the potential and magnetic field:

\begin{equation}
\nabla p = -\rho \nabla \phi + \mathbf{J} \times \mathbf{B},
\end{equation}
where $p = \rho c_s^2$ is the thermal pressure with the isothermal sound speed $c_s$, $\phi$ is the gravitational potential, $\mathbf{J} = \nabla \times \mathbf{B} / 4\pi$ is the current density and $\mathbf{J} \times \mathbf{B}$ is the Lorentz force density. This yields
\begin{equation}
\nabla \ln \rho = -\nabla \phi / c_s^2 + (\mathbf{J} \times \mathbf{B}) / (\rho c_s^2).
\end{equation}

The initial density distribution is first solved out in cylindrical grids and then mapped into our cartesian grids for the simulations. The cylindrical density distribution is obtained in two steps using a fourth-order Runge--Kutta integration scheme. First, the density profile of the midplane $\rho(R, z=0)$ is calculated by integrating radially inward from the outer boundary $R_{\rm out}=100$ kpc:
\begin{equation}
\mathrm{d} \ln \rho / \mathrm{d} R = -\left( \partial \phi / \partial R \right) / c_s^2.
\end{equation}
The gas density at the outer boundary $R_{\rm out}$ is determined from the approximate distribution $n = 0.02 \, (r / \mathrm{kpc})^{-1.5}$ cm$^{-3}$ from our previous study \cite{Zhang2021}. Then, for each fixed $R$, the vertical profile $\rho(R, z)$ in the $z$ direction is integrated outward from the midplane:
\begin{equation}
\mathrm{d} \ln \rho / \mathrm{d} z = \left( -\partial \phi / \partial z + (\mathbf{J} \times \mathbf{B})_z / \rho \right) / c_s^2.
\end{equation}

\subsubsection*{Magnetic Field Configuration}

The initial magnetic field is divergence-free and axisymmetric in cylindrical coordinates $(R, \phi, z)$, comprising toroidal and poloidal components.

The toroidal component decreases exponentially with height from the Galactic disk \cite{Unger2024}, with an inner suppression term $\sigma\left(\frac{R - r_1}{w_1}\right)$ to account for turbulence from the Galactic bar, star formation, and AGN feedback in the central region:
\begin{equation}
B_\phi(R, z) = B_0 \left[1 - h_d(z)\right] e^{-\vert z\vert/z_t} \left[1 - \sigma\left(\frac{R - r_t}{w_t}\right)\right] \sigma\left(\frac{R - r_1}{w_1}\right),
\end{equation}
where $B_0 = 3.0 \, \mu$G. The vertical fade-in function $h_d(z)$ is defined as:
\begin{equation}
  h_d(z) = 1 - \sigma\left( \frac{\vert z\vert - z_d}{w_d} \right) , 
\end{equation}
with $z_d = 0.3$ kpc, $w_d = 0.3$ kpc, $z_t = 4.0$ kpc, $r_t = 10$ kpc, $w_t = 1.7$ kpc, $r_1 = 5.0$ kpc, and $w_1 = 4.0$ kpc. The sigmoid function is $\sigma(x) = 1 / (1 + e^{-x})$.

The poloidal component follows an X-shaped form \cite{Ferriere2014,Unger2024}:
\begin{eqnarray}
B_z(R, z) &=& B_1 \left( \frac{r_1^2}{R^2} \right) \exp\left( -\frac{r_1}{L} \right), \\
B_R(R, z) &=& B_1 \left( \frac{2 a z r_1^3}{R^2} \right) \exp\left( -\frac{r_1}{L} \right),
\end{eqnarray}
where $r_1 = R / (1 + a z^2)$, $B_1$ =1.0 $\mu$G, $a = 0.01$ kpc$^{-2}$, and $L = 10.0$ kpc.

\begin{figure}[htbp]
	\centering
	\begin{subfigure}{0.45\linewidth}
		\centering
		\includegraphics[width=0.9\linewidth]{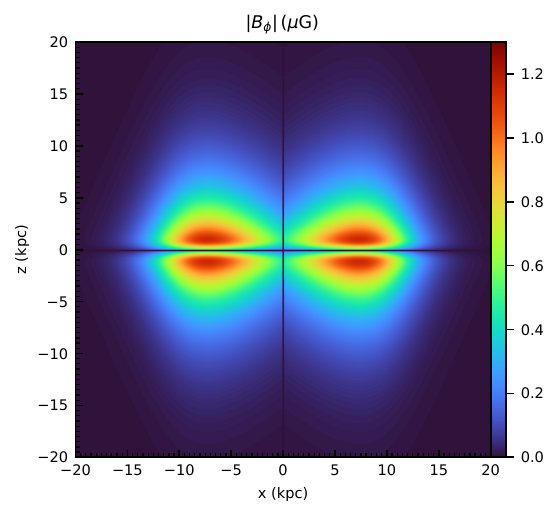}
		\caption{Toroidal component.}
		\label{B_tor}
	\end{subfigure}
	\begin{subfigure}{0.45\linewidth}
		\centering
		\includegraphics[width=0.9\linewidth]{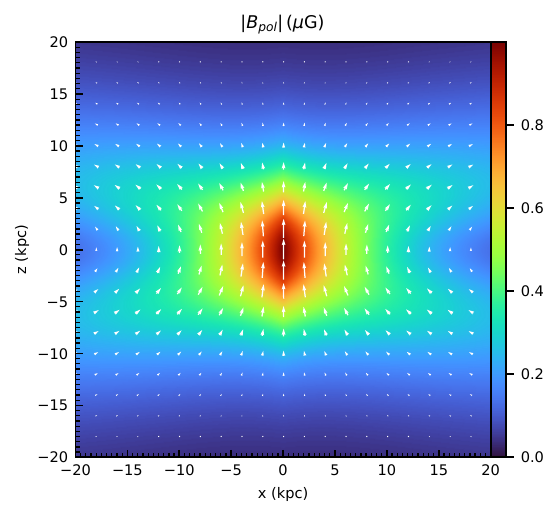}
		\caption{Poloidal component.}
		\label{B_pol}
	\end{subfigure}
	\caption{Central slices of the initial magnetic field strength distribution (Left: toroidal component; Right: poloidal component) in the $x$--$z$ plane. White arrows in the right panel show the direction of the poloidal magnetic field.}
	\label{B0}
\end{figure}

The above magnetic field components are then mapped to our Cartesian grid. The initial field slices are shown in Fig.~\ref{B0}.

\subsection*{Jet Injection}

A pair of opposing jets are injected into a compact cuboidal region with a base side length $L_{\rm jet}=15.6$ pc and a height $z_{\rm jet}=350$ pc per side, with a total volume $V_{\rm jet}$. The interval between episodes and the internal-to-kinetic energy ratio are free parameters, chosen to match the observed bubble properties at $t=15$ Myr. In Supplementary Information, we discuss how some key jet parameters are constrained by current observations.

An energy-conserving scheme injects mass, momentum, and energy into the cuboid. The injection rate of the internal energy density is
\begin{equation}
\dot{e}_{\rm internal} = \frac{E_{\rm internal}}{V_{\rm jet} t_{\rm jet}}.
\end{equation}
The density injection rate is
\begin{equation}
\dot{\rho}_{\rm jet} = \frac{2 E_{\rm kinetic}}{V_{\rm jet} v_{\rm jet}^2 t_{\rm jet}}.
\end{equation}
Cell quantities are updated as:
\begin{align}
\rho_{\rm new} &= \rho_{\rm old} + \dot{\rho}_{\rm jet} \Delta t, \\
\mathbf{v}_{\rm new} &= \frac{\rho_{\rm old} \mathbf{v}_{\rm old} + \dot{\rho}_{\rm jet} \Delta t \, \mathbf{v}_{\rm jet}}{\rho_{\rm new}}, \\
p_{\rm new} &= (\gamma - 1) e_{\rm internal, new},
\end{align}
where
\begin{align}
    &e_\mathrm{internal,new} = e_\mathrm{kinetic,old} + e_\mathrm{internal,old} + e_\mathrm{kinetic,in} + e_\mathrm{internal,in} - e_\mathrm{kinetic,new},\\
    &e_\mathrm{kinetic,old} = 0.5 \rho_\mathrm{old} v_\mathrm{old}^2, \\
    &e_\mathrm{internal,old} = p_\mathrm{old} / (\gamma - 1), \\
    &e_\mathrm{kinetic,in} = 0.5 \dot{\rho}_\mathrm{jet} \Delta t \, v_{jet}^2, \\
    &e_\mathrm{internal,in} = \dot{e}_\mathrm{internal} \Delta t, \\
    &e_\mathrm{kinetic,new} = 0.5 \rho_\mathrm{new} v_\mathrm{new}^2.
\end{align}

\subsection*{Post-Processing of Multi-Wavelength Emissions}

Synthetic maps for X-ray, gamma-ray, and synchrotron emissions are generated from the simulation output to compare with observations. All integrations are performed along lines of sight from the Sun's position, assuming optically thin emission unless noted otherwise.

\subsubsection*{Synthetic X-ray Maps and Line Ratios}

X-ray surface brightness maps are computed by integrating thermal emissivities in the ROSAT bands (0.11--0.284 keV, 0.44--1.21 keV, 0.73--2.04 keV) and eROSITA equivalents. The surface brightness $I(l, b)$ in Galactic coordinates is
\begin{equation}
I(l, b) = \frac{1}{4\pi} \int_{\rm los} n_{\rm H} n_{\rm e} \epsilon(T, Z) \, dl \quad (\rm erg \, s^{-1} \, cm^{-2} \, sr^{-1}),
\end{equation}
where $n_{\rm H}$ and $n_{\rm e}$ are hydrogen and electron number densities, and $\epsilon(T, Z)$ is the emissivity from the APEC model \cite{Smith2001} via {\it pyatomdb}, including line and bremsstrahlung emissions. Uniform metallicity $Z = 0.3 Z_\odot$ is assumed for the hot plasma. Integration extends to a distance of 80 kpc.

Photoelectric absorption by neutral hydrogen is applied using
\begin{equation}
I'(l, b) = I(l, b) e^{-\sigma N_{\rm HI}},
\end{equation}
where $N_{\rm HI}$ is adopted from the Effelsberg-Bonn HI Survey \cite{Winkel2016} and $\sigma$ is the cross-section \cite{Wilms2000}.  To account for foreground emission in 0.11-0.284 keV, we added a uniform surface brightness of 1 cts s$^{-1}$ deg$^{-2}$ to the synthetic map, a value close to ROSAT measurements in the optically thick Galactic plane within this band. The effective areas of ROSAT at 0.11-0.284 keV, 0.44-1.21 keV, 0.73-2.04 keV are approximated to be 200 cm$^2$, 100 cm$^2$, and 200 cm$^2$, respectively.

The surface brightness distributions of the O VII and O VIII lines are similarly computed, incorporating absorption, for direct comparison with the eROSITA line-ratio maps. 

\subsubsection*{Gamma-ray Emission}

\begin{figure}[htbp]
\centering
\includegraphics[width=0.9\linewidth]{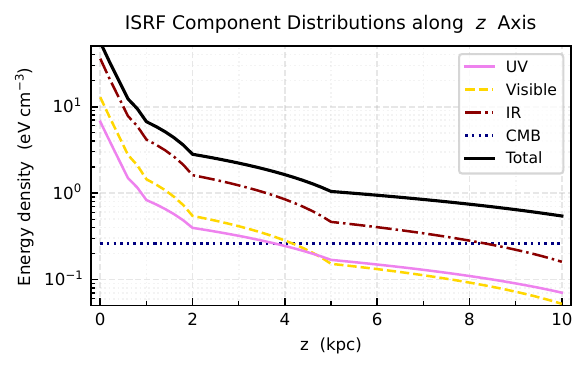}
\caption{Energy density variations of the ISRF components along the \( z \) axis (from GALPROP).}
\label{ISRF}
\end{figure}

Gamma-ray emission is modeled through inverse Compton scattering (ICS) of relativistic electrons off the interstellar radiation field (ISRF), including the contributions from the cosmic microwave background (CMB), infrared (IR), visible and ultraviolet (UV) components, adopted from the GALPROP model \cite{Porter2017}. The CMB is treated as a blackbody spectrum with temperature \( T = 2.725 \,\mathrm{K} \) and uniform energy density \( u_{\rm CMB} = 0.26 \,\mathrm{eV\,cm^{-3}} \). The spatial variations of the ISRF components along the \( z \) axis are shown in Fig.~\ref{ISRF}.

We assume that the non-thermal gamma ray-emitting electron energy density is proportional to the thermal internal energy density of the hot gas,
\begin{equation}
u_{\rm e} = f_{\rm IC}\, u_{\rm internal},
\end{equation}
where \( u_{\rm internal} \) is obtained from our MHD simulations, and \( f_{\rm IC} = 2.5 \times 10^{-3} \) is calibrated to match the observed gamma-ray intensity. The CRe spectral distribution is modeled as a power law in the Lorentz factor ($\gamma$) with an exponential cutoff \cite{Ackermann2014}:
\begin{equation}
\frac{dN_e}{d\gamma} = C_r \, \gamma^{-p} \, \exp\!\left(-\frac{\gamma}{\gamma_{\rm cut}}\right),
\end{equation}
where \( p \) is the spectral index. We adopt \( \gamma_{\rm cut} = 2.5 \times 10^6 \) (corresponding to \( \sim 1.3 \)~TeV) and \( \gamma_{\rm min} = 2000 \) (\( \sim 1 \)~GeV). The spectral index varies spatially, with \( p = 2.2 \) inside the Fermi bubbles and \( p = 2.4 \) outside, consistent with observational constraints \cite{Ackermann2014}. The normalization constant \( C_r \) is determined by the CRe energy density:
\begin{equation}
C_r = \frac{u_{\rm e}}{m_e c^2 \int_{\gamma_{\rm min}}^{\infty} \gamma^{1-p} \exp(-\gamma/\gamma_{\rm cut}) \, d\gamma},
\end{equation}
where $c$ is the speed of light.  

The differential gamma-ray emissivity \( j(E_\gamma) \) from ICS is computed as
\begin{equation}
j(E_\gamma) =  \frac{1}{4\pi}\int \frac{dn}{d\varepsilon}
\left[
\int_{\gamma_{\rm min}}^{\infty}
c \, \frac{d\sigma_{\rm KN}}{dE_\gamma} \,
\frac{dN_e}{d\gamma} \, d\gamma
\right] d\varepsilon,
\end{equation}
where $j(E_\gamma)$ is in unit of (GeV$^{-1}$cm$^{-2}$s$  ^{-1}$), \( \varepsilon \) denotes the photon energy of the ISRF,
\( dn/d\varepsilon \) is the corresponding differential photon number density,
\( E_\gamma \) is the energy of the upscattered gamma-ray photons, and
the scatting cross section is \cite{Blumenthal1970}:
\begin{align}
\frac{d\sigma_{\rm IC}}{dE_\gamma} &= \frac{3\sigma_T}{4 \gamma^2 \varepsilon}
\left[
2q\ln q + (1+2q)(1-q)
+ \frac{1}{2}(\Gamma_\varepsilon q)^2
\frac{1-q}{1+\Gamma_\varepsilon q}
\right], \\
\Gamma_\varepsilon &= \frac{4\varepsilon \gamma}{m_e c^2}, \qquad
q = \frac{E_\gamma}{\Gamma_\varepsilon (\gamma m_e c^2 - E_\gamma)},
\end{align}
with the Thomson cross section \( \sigma_T = 6.65 \times 10^{-25}\,\mathrm{cm}^2 \).

The observed gamma-ray surface brightness is obtained by integrating the emissivity along the line of sight up to a distance of $30$ kpc:
\begin{equation}
I(E_\gamma) =\int_{\rm los} j(E_\gamma) \, ds,
\end{equation}
where \( s \) is the line-of-sight distance.

\subsubsection*{Synchrotron Emission}

We compute the Galactic synchrotron emission using the outputs of the three-dimensional distributions of magnetic field strength and hot gas properties obtained from our MHD simulations, following the synchrotron formalism implemented in the {\it hammurabi X} code \cite{Ginzburg1965,Wang2020}. Both total and polarized synchrotron emissions are calculated. Faraday rotation, synchrotron self-absorption, and free--free absorption are neglected because they are subdominant at the frequencies considered. We adopt the IAU polarization convention, in which polarization angles are measured as positive from the north toward the east.

The synchrotron-emitting electrons are assumed to follow a power-law energy distribution,
\begin{equation}
\frac{dN_e}{d\gamma} = N_0\, \gamma^{-p},
\end{equation}
where \( \gamma \) is the electron Lorentz factor, \( p \) is the spectral index, and \( N_0 \) is the normalization constant. Observations indicate that the electron spectral index varies significantly across different regions of the bubbles, ranging from \( p \sim 1.8 \) in the northern Fermi bubble to \( p \sim 3 \) in the NPS, and also depends on the observation frequency \cite{Guidi2023}. For consistency with the gamma-ray modeling, we adopt the same spectral indices, with \( p = 2.2 \) inside the Fermi bubbles and \( p = 2.4 \) outside.

The energy density of relativistic electrons may then be calculated by
\begin{equation}
u_e = \int_{\gamma_{\min}}^{\gamma_{\max}} \gamma m_e c^2 \frac{dN_e}{d\gamma}\, d\gamma,
\end{equation}
where \( \gamma_{\min} \) and \( \gamma_{\max} \) denote the minimum and maximum Lorentz factors of the synchrotron-emitting electrons, respectively. The normalization constant \( N_0 \) is then determined as \cite{Duan2025}:
\begin{equation}
N_0 =
\begin{cases}
\dfrac{u_e}{m_e c^2} \ln\!\left(\dfrac{\gamma_{\max}}{\gamma_{\min}}\right), & p = 2,\\[10pt]
\dfrac{u_e}{m_e c^2} \dfrac{2-p}{\gamma_{\max}^{2-p}-\gamma_{\min}^{2-p}}, & p \neq 2.
\end{cases}
\label{eq:N0_syn}
\end{equation}
We adopt \( \gamma_{\min} = 100 \) and \( \gamma_{\max} = 10^{5} \), consistent with the typical values inferred for Galactic synchrotron-emitting electrons \cite{Hardcastle2018}.

We further assume that the energy density of synchrotron-emitting electrons scales with the local gas internal energy density,
\begin{equation}
u_{\rm e} = f_{\rm syn} u_{\rm internal},
\end{equation}
where \( u_{\rm internal} \) is the thermal internal energy density obtained from our simulations. The normalization factor is set to \( f_{\rm syn} = 0.025 \), chosen to best reproduce the observed large-scale synchrotron surface brightness of the bubbles. Note that \( f_{\rm syn} \) is treated independently from the normalization adopted for inverse-Compton emission, allowing for different effective electron populations responsible for synchrotron and gamma-ray radiations.

For an observer located at $(-8.3,\,0,\,0)\,\mathrm{kpc}$, the line-of-sight unit vector toward Galactic longitude $l$ and latitude $b$ is given by
\begin{equation}
\hat{u} = (\cos b \cos l,\, \cos b \sin l,\, \sin b),
\end{equation}
with the corresponding north and east basis vectors
\begin{equation}
\hat{e}_{\rm north} = (-\sin b \cos l, -\sin b \sin l, \cos b), \quad
\hat{e}_{\rm east} = (-\sin l, \cos l, 0).
\end{equation}

Along the line of sight parameterized by distance $s$, the magnetic field component perpendicular to the line of sight is
\begin{equation}
\mathbf{B}_\perp = \mathbf{B} - (\mathbf{B}\cdot\hat{u})\hat{u}.
\end{equation}

The total synchrotron emissivity is given by \cite{Wang2020}
\begin{equation}
j_I =
\frac{\sqrt{3} e^3 B_\perp N_0}{4\pi m_e c^2 (p + 1)}
\left( \frac{2\pi \nu m_e c}{3 e B_\perp} \right)^{(1 - p)/2}
\Gamma\!\left( \frac{p}{4} + \frac{19}{12} \right)
\Gamma\!\left( \frac{p}{4} - \frac{1}{12} \right),
\end{equation}
while the polarized emissivity is
\begin{equation}
j_{P} =
\frac{\sqrt{3} e^3 B_\perp N_0}{16\pi m_e c^2}
\left( \frac{2\pi \nu m_e c}{3 e B_\perp} \right)^{(1 - p)/2}
\Gamma\!\left( \frac{p}{4} + \frac{7}{12} \right)
\Gamma\!\left( \frac{p}{4} - \frac{1}{12} \right).
\end{equation}

The intrinsic magnetic position angle on the plane of the sky is defined as
\begin{equation}
\psi = \rm{atan2}(\mathbf{B}_\perp \cdot \hat{e}_{\rm east},
               \mathbf{B}_\perp \cdot \hat{e}_{\rm north}),
\end{equation}
where the $\rm{atan2}$ function represents the quadrant-preserving arc tangent. The corresponding intrinsic Stokes parameters are
\begin{equation}
q = -j_P \cos 2\psi, \qquad
u = j_P \sin 2\psi.
\end{equation}
following the IAU polarization convention. 

The observed Stokes parameters are obtained by integrating the emissivities along the line of sight up to a maximum distance of $30\,\mathrm{kpc}$,
\begin{equation}
I = \int_{\rm LoS} j_I(s)\, ds, \qquad
Q = \int_{\rm LoS} q(s)\, ds, \qquad
U = \int_{\rm LoS} u(s)\, ds,
\end{equation}
and the polarized intensity is defined as $PI = \sqrt{Q^2 + U^2}$.

\section*{Supplementary Information}
\subsection*{Parameter Dependence and Constraints}\label{paraserv}

Given the current observational constraints, the detailed jet parameters cannot be uniquely determined, and multiple combinations may reproduce broadly similar large-scale morphologies. Nevertheless, several key physical quantities are robustly constrained across the parameter space explored. Despite degeneracies in individual jet parameters, the global energetics, the temporal sequence, and the dual-shock structure of the bubbles are robustly determined.

\begin{figure}[htbp]
\centering
\begin{subfigure}{0.49\textwidth}
\centering
\includegraphics[width=\linewidth]{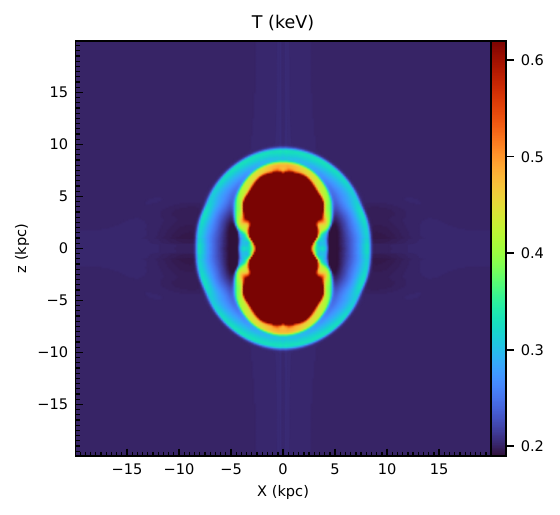}
\centering
\caption{}
\label{modelB}
\end{subfigure}\hfill
\begin{subfigure}{0.49\textwidth}
\centering
\includegraphics[width=\linewidth]{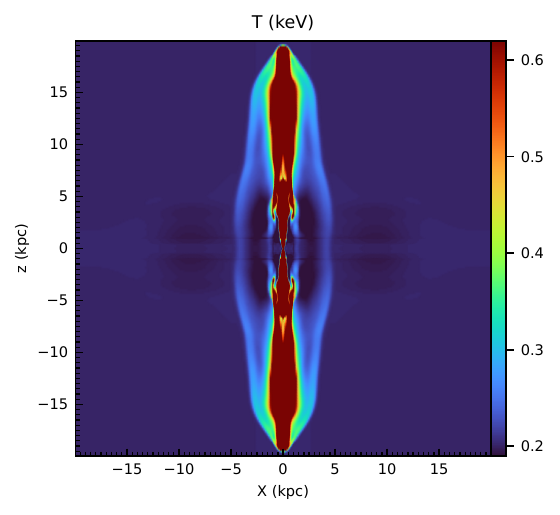}
\centering
\caption{}
\label{modelC}
\end{subfigure}
\caption{Temperature distributions in parameter-varied simulation runs (A, B) at t = 15 Myr, illustrating the dependence of the bubble morphology on key jet parameters relative to the fiducial model (see Table~\ref{tab:jet_parameters}). Run A (left) is dominated by internal energy, while run B (right) features a longer injection duration. A double shock structure is present in both runs. }
\label{fig:parameter_models}  
\end{figure}

First, the total injected energy per jet episode—along with the partitioning between kinetic and internal energy—is reasonably constrained by the observations of the eROSITA and Fermi bubbles. The overall energy budget primarily determines the downstream gas temperatures at the current locations of the propagating forward shocks (bubble edges), whereas a dominance of kinetic over internal energy is required to reproduce their observed elongation along the Galactic rotation axis. Models dominated by internal energy injection tend to produce overly spherical morphologies  (Fig.~\ref{modelB}), inconsistent with the shapes of both the eROSITA and Fermi bubbles.

Second, the ages of the eROSITA and Fermi bubbles are among the most tightly constrained quantities. The age of the Fermi bubbles is constrained by the Mach numbers inferred at their edges ($\mathcal{M}\sim1.5$--2 \cite{Kataoka2015,Miller2016}, corresponding to a dynamical age of $\sim$5--6 Myr \cite{Zhang2020}). In contrast, the age of the eROSITA bubbles is independently constrained by the observed O VIII/O VII line ratios in the eROSITA narrow-band maps, which are sensitive to the post-shock gas temperature and the collisional ionization timescale. Self-consistently reproducing both the morphology and the X-ray line ratios of the eROSITA bubbles requires a distinct, older shock driven by an earlier jet episode.

Third, the vertical extents (heights) of the bubble pairs offer an additional geometric constraint. For jets aligned with the Galactic rotation axis, the observed heights and bubble morphology directly limit the durations of the two jet episodes. A jet that is too long-lived produces narrow, under-expanded bubble tops that are inconsistent with observations (Fig.~\ref{modelC}). Even if the jets are moderately inclined relative to the Galactic plane—making the intrinsic physical heights larger than their projected extents—the observed relative separation between the eROSITA and Fermi bubble edges remains a robust diagnostic.

However, some other parameters exhibit partial degeneracies. While the jet power and energy are constrained, the jet speed $v_{\rm jet}$, the jet mass density, and the jet base size $R_{\rm jet}$ are somewhat degenerate from each other in our energy-conserving injection scheme. Importantly, regardless of these degeneracies, a single jet episode is fundamentally more difficult to reconcile with the distinct spatial extents, ages, and multi-wavelength properties of the eROSITA and Fermi bubbles. Across our entire parameter space, only models featuring two temporally separated jet episodes successfully generate the successive forward shocks needed to account for the observed spatial stratification in X-ray, radio, and gamma-ray emissions. This conclusion is robust, emerging independently of specific jet configurations and without requiring fine-tuning of either jet parameters or the halo gas profile.

\begin{table}[htbp]

\centering
\caption{Simulation parameters adopted for the jet episode. Run A is dominated by internal energy ($E_{\rm internal} \gg E_{\rm kinetic}$); Run B features a longer injection duration ($t_{\rm jet}$); and Run ``Fiducial'' is the fiducial model presented in the main text. In all simulations, the second jet is injected at $t = 10$ Myr, with $v_{\rm jet2} = 0.65 v_{\rm jet1}$ and $E_{\rm kinetic2} = 0.27E_{\rm kinetic1}$, while all other independent jet parameters remain the same as those of the first episode. Here, the subscript $1$ refers to the first jet, and $2$ to the second jet.  }
\label{tab:jet_parameters}
\begin{tabular}{lcccccc}
\toprule
Run & $E_{\rm total1}$ & $E_{\rm kinetic1}$ & $E_{\rm internal1}$ & $t_{\rm jet}$ & $v_{\rm jet1}$ & $E_{\rm total2}$ \\
    & (erg)                & (erg)                & (erg)                & (Myr)        & (cm s$^{-1}$)      & (erg)              \\
\midrule
A & $5.65\times10^{55}$ & $5.20\times10^{54}$ & $5.13\times10^{55}$ & 1.0 & $1.00\times10^{9}$ & $5.27\times10^{55}$\\
B & $7.19\times10^{54}$ & $5.04\times10^{54}$ & $2.05\times10^{54}$ & 8.0 & $0.50\times10^{9}$ & $3.47\times10^{54}$\\
Fiducial & $3.46\times10^{55}$ & $3.25\times10^{55}$ & $2.11\times10^{54}$ & 1.0 & $1.85\times10^{9}$ & $1.10\times10^{55}$\\
\bottomrule
\end{tabular}
\end{table}

\section*{Data availability}
This paper makes use of the ROSAT data, publicly available through the MPE and HEASARC archives at GSFC.  The EBHIS data sets are available  via
\url{http://cdsweb.u-strasbg.fr/cgi-bin/qcat?J/A+A/585/A41}. Atom emissivity data are available at \url{http://www.atomdb.org/index.php}. Synchrotron polarization maps are available at \url{https://portal.nersc.gov/project/cmb/Planck_Revisited/}
The simulation data supporting the findings of this study are available on request from the corresponding author.

\section*{Code Availability}
The simulations were performed using the code PLUTO, publicly available at \url{https://plutocode.ph.unito.it/}. The X-ray emissivity data is extracted with pyatomdb,publicly available at \url{https://atomdb.readthedocs.io/en/master/}.

\backmatter

\section*{Acknowledgements}
{
The authors would like to thank the anonymous reviewers for their valuable comments and suggestions that greatly strengthened the manuscript. Ruiyu Zhang thanks Li Ji, Xueying Zheng and Peter Predehl for helpful discussions. We acknowledge the use of the data from the ROSAT mission (a joint German-US-UK project supported by BMFT, NASA, and MPG), the Effelsberg-Bonn HI Survey (EBHIS, obtained with the 100-m telescope at Effelsberg operated by MPIfR), AtomDB for atomic data and plasma modeling, and low-noise synchrotron polarization maps from NASA LAMBDA (HEASARC), including Planck (ESA mission with contributions from ESA Member States, NASA, and Canada) and WMAP (supported by NASA Office of Space Sciences). This research used healpy, matplotlib, NumPy, and Astropy. This work was supported by the National SKA Program of China (No. 2025SKA0130100), the China Postdoctoral Science foundation (No. 2023M731014), the Excellent Youth Team Project of the Chinese Academy of Sciences (No. YSBR-061), the National Natural Science Foundation of China (No. 12473010), Shanghai Pilot Program for Basic Research - Chinese Academy of Sciences, Shanghai Branch (No. JCYJ-SHFY-2021-013), and the China Manned Space Program (No. CMS-CSST-2025-A10, CMS-CSST-2025-A08). H.S. was  supported by the National Natural Science Foundation of China (Nos. 11863003, 12173010, 12573034). G.M. was supported by the NSFC (Nos. 12473013, 12133007). The simulations were performed in the High performance computing center of Henan Normal University.
} 

\section*{Author Contributions}
{
Ruiyu Zhang conceived the idea, designed and performed the simulations, and drafted the manuscript. Fulai Guo supervised the project, contributed to the interpretation of the results, and played a leading role in significantly reworking and extensively revising the manuscript. Shaokun Xie and Ruofei Zhang developed the initial conditions and parameters for the simulation code. Hanfeng Song and Shumin Wang conducted data analysis and generated synthetic X-ray maps. Guobin Mou revised and provided critical feedback on the manuscript. Xiaodong Duan provided critical feedback on the manuscript and contributed to the setup of the magnetic field  during the revision. All authors reviewed and approved the final manuscript. 
}
\section*{Competing interests} {The authors declare no competing interests.}

\newcommand\aap{A\&A}                
\let\astap=\aap                          
\newcommand\aapr{A\&ARv}             
\newcommand\aaps{A\&AS}              
\newcommand\actaa{Acta Astron.}      
\newcommand\afz{Afz}                 
\newcommand\aj{AJ}                   
\newcommand\ao{Appl. Opt.}           
\let\applopt=\ao                         
\newcommand\aplett{Astrophys.~Lett.} 
\newcommand\apj{ApJ}                 
\newcommand\apjl{ApJ}                
\let\apjlett=\apjl                       
\newcommand\apjs{ApJS}               
\let\apjsupp=\apjs                       
\newcommand\apss{Ap\&SS}             
\newcommand\araa{ARA\&A}             
\newcommand\arep{Astron. Rep.}       
\newcommand\aspc{ASP Conf. Ser.}     
\newcommand\azh{Azh}                 
\newcommand\baas{BAAS}               
\newcommand\bac{Bull. Astron. Inst. Czechoslovakia} 
\newcommand\bain{Bull. Astron. Inst. Netherlands} 
\newcommand\caa{Chinese Astron. Astrophys.} 
\newcommand\cjaa{Chinese J.~Astron. Astrophys.} 
\newcommand\fcp{Fundamentals Cosmic Phys.}  
\newcommand\gca{Geochimica Cosmochimica Acta}   
\newcommand\grl{Geophys. Res. Lett.} 
\newcommand\iaucirc{IAU~Circ.}       
\newcommand\icarus{Icarus}           
\newcommand\japa{J.~Astrophys. Astron.} 
\newcommand\jcap{J.~Cosmology Astropart. Phys.} 
\newcommand\jcp{J.~Chem.~Phys.}      
\newcommand\jgr{J.~Geophys.~Res.}    
\newcommand\jqsrt{J.~Quant. Spectrosc. Radiative Transfer} 
\newcommand\jrasc{J.~R.~Astron. Soc. Canada} 
\newcommand\memras{Mem.~RAS}         
\newcommand\memsai{Mem. Soc. Astron. Italiana} 
\newcommand\mnassa{MNASSA}           
\newcommand\mnras{MNRAS}             
\newcommand\na{New~Astron.}          
\newcommand\nar{New~Astron.~Rev.}    
\newcommand\nat{Nature}              
\newcommand\nphysa{Nuclear Phys.~A}  
\newcommand\pra{Phys. Rev.~A}        
\newcommand\prb{Phys. Rev.~B}        
\newcommand\prc{Phys. Rev.~C}        
\newcommand\prd{Phys. Rev.~D}        
\newcommand\pre{Phys. Rev.~E}        
\newcommand\prl{Phys. Rev.~Lett.}    
\newcommand\pasa{Publ. Astron. Soc. Australia}  
\newcommand\pasp{PASP}               
\newcommand\pasj{PASJ}               
\newcommand\physrep{Phys.~Rep.}      
\newcommand\physscr{Phys.~Scr.}      
\newcommand\planss{Planet. Space~Sci.} 
\newcommand\procspie{Proc.~SPIE}     
\newcommand\rmxaa{Rev. Mex. Astron. Astrofis.} 
\newcommand\qjras{QJRAS}             
\newcommand\sci{Science}             
\newcommand\skytel{Sky \& Telesc.}   
\newcommand\solphys{Sol.~Phys.}      
\newcommand\sovast{Soviet~Ast.}      
\newcommand\ssr{Space Sci. Rev.}     
\newcommand\zap{Z.~Astrophys.}       

\bibliography{ancillary/astrobib}

\end{document}